\documentclass[conference]{IEEEtran}
\usepackage{cite}
\usepackage{amsmath,amssymb,amsfonts}
\usepackage{algorithmic}
\usepackage{graphicx}
\usepackage{siunitx}
\usepackage[bookmarks=false]{hyperref}
\usepackage{textcomp}
\usepackage{xcolor}
\def\BibTeX{{\rm B\kern-.05em{\sc i\kern-.025em b}\kern-.08em
    T\kern-.1667em\lower.7ex\hbox{E}\kern-.125emX}}
    
\makeatletter
\newcommand*\titleheader[1]{\gdef\@titleheader{#1}}
\AtBeginDocument{%
  \let\st@red@title\@title
  \def\@title{%
    \bgroup\normalfont\large\centering\@titleheader\par\egroup
    \vskip0.5em\st@red@title}
}
\makeatother

\title{\huge Analysis of Contraction Effort Level in EMG-Based Gesture Recognition Using Hyperdimensional Computing
}
\titleheader{\vspace{-30pt}Published as a conference paper at the IEEE BioCAS 2019}

\author{\IEEEauthorblockN{Ali Moin\IEEEauthorrefmark{1}, Andy Zhou\IEEEauthorrefmark{1}, Simone Benatti\IEEEauthorrefmark{2}, Abbas Rahimi\IEEEauthorrefmark{1}\IEEEauthorrefmark{3}, Luca Benini\IEEEauthorrefmark{2}\IEEEauthorrefmark{3}, Jan M. Rabaey\IEEEauthorrefmark{1}}
\IEEEauthorblockA{\IEEEauthorrefmark{1}Berkeley Wireless Research Center, EECS Department, 
University of California, Berkeley.}
\IEEEauthorblockA{\IEEEauthorrefmark{2}DEI, University of Bologna, Italy. \IEEEauthorrefmark{3}Integrated System Laboratory, ETH Zurich, Switzerland.}
\IEEEauthorblockA{Corresponding Author Email: moin@berkeley.edu}
}

\begin{document}

\maketitle

\begin{abstract}
Varying contraction levels of muscles is a big challenge in electromyography-based gesture recognition. Some use cases require the classifier to be robust against varying force changes, while others demand to distinguish between different effort levels of performing the same gesture. We use brain-inspired hyperdimensional computing paradigm to build classification models that are both robust to these variations and able to recognize multiple contraction levels. Experimental results on 5 subjects performing 9 gestures with 3 effort levels show up to 39.17\% accuracy drop when training and testing across different effort levels, with up to 30.35\% recovery after applying our algorithm.

\end{abstract}



\section{Introduction}
Hand gestures are an integral part of human communication as well as object manipulation and dexterity. Electromyography (EMG)-based pattern recognition has shown great potential in classifying hand gestures, where EMG features gathered from the sensors on the skin serve as inputs to machine learning algorithms. Although being non-invasive makes it an attractive method, it is highly prone to signal variations caused by factors such as changing limb position~\cite{geng2012toward}, electrode shift~\cite{young2011improving}, and force change~\cite{al2015improving}. While the first two are undesired phenomena that the classifier has to ideally be robust against, the last could occasionally be desired in applications such as proportional control of prosthetic hands.

A subject can exert different levels of effort while performing a gesture, resulting in different EMG signal properties. Scheme and Englehart~\cite{scheme2011electromyogram} have shown up to 50\% error rate when the classifier was trained and tested at different force levels from 20\% to 80\% maximal voluntary contraction (MVC), compared to moderate 7\% to 19\% error rate when trained and tested at the same level. Previous works have suggested to pick specific force levels that yield minimum accuracy degradation across all force levels as training dataset and to extract features that are  more invariant against contraction levels as the input to the classifier~\cite{khushaba2016combined,he2014invariant}.

\begin{figure}
	\centering
	\includegraphics[width=0.55\columnwidth]{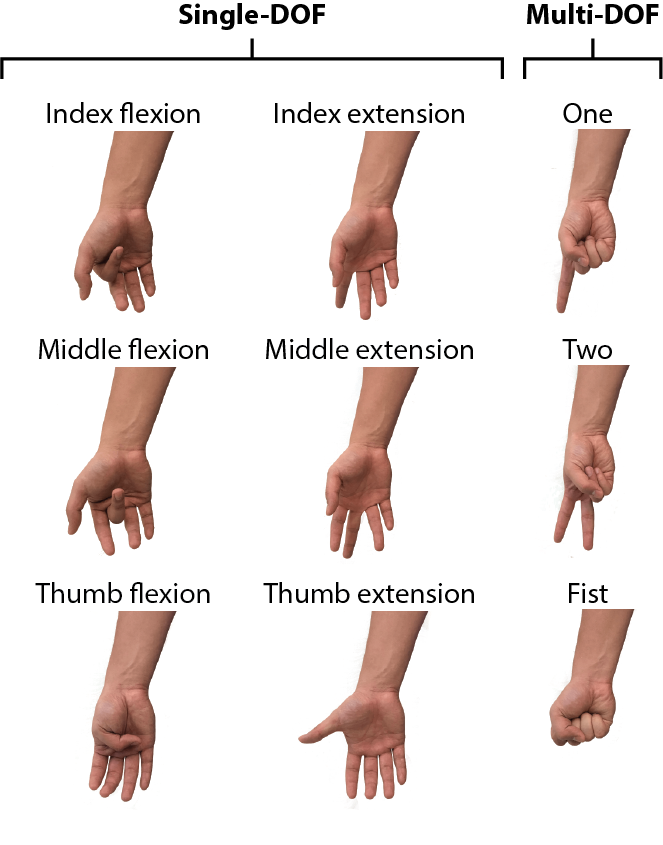}
	\vspace{-15pt}
	\caption{Hand gesture classes used in the study. The single degree-of-freedom (DOF) gesture subset includes individual finger flexions and extensions. The multi-DOF gesture subset includes isometric hand postures involving multiple fingers.}
	\label{fig:gests}
\end{figure}

In this paper, we propose building a general classification model based on hyperdimensional (HD) computing~\cite{kanerva2009hyperdimensional} to deal with varying muscle contraction effort levels. HD computing has shown promising results in classification tasks using biosignals such as EMG in recognizing hand gestures~\cite{moin2018emg} and electrocorticography (ECoG) for seizure detection with one-shot learning~\cite{burrello2018one}. With slight modifications to our previously introduced encoding scheme~\cite{moin2018emg}, we analyze the muscle contraction level variations in two different ways, depending on the application: If discrimination between different gestures is the only goal, the classifier should output the same gesture class regardless of the subject's effort level. If, on the other hand, different effort levels are relevant to the application (e.g. controlling different levels of force for gripping using a prosthetic hand), different effort levels for the same gesture must be treated as separate output classes. A classifier based on HD computing can be naturally used in both of these scenarios. In the former case, it can include minimum amount of training data from multiple effort levels for training each gesture to build an inclusive model that ignores effort level variations. In the latter, distinguishing between different effort levels of the same gesture translates to simply defining a separate class for each level of contraction.
A dataset of 5 human subjects performing 9 hand gestures (Fig.~\ref{fig:gests}) with low, medium, and high contraction effort levels was recorded using a wireless, high-channel count EMG recording system~\cite{moin2018emg} which provided visual feedback of effort level. Classification accuracy results for both gesture-only and gesture+effort cases are presented.


\section{Experiment Setup}

\begin{figure}
	\centering
	\includegraphics[width=\columnwidth]{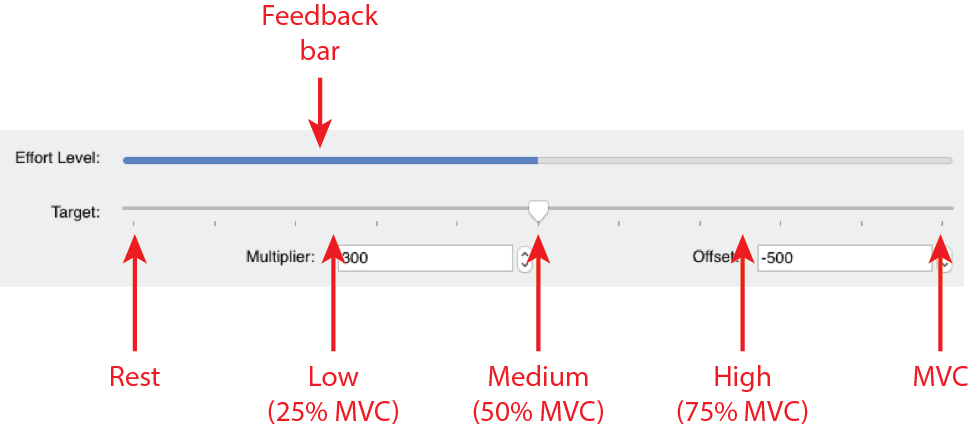}
	\vspace{-20pt}
	\caption{Visual feedback of real-time contraction effort level to the user. The bar represents the mean signal energy across all channels as a measure of contraction effort. The users are asked to reach 25\%, 50\%, and 75\% of their maximum voluntary contraction (MVC). During calibration, a multiplier and an offset are determined such that the rest state and the MVC map to 0 and 100, respectively.}
	\label{fig:bar}
	\vspace{-10pt}
\end{figure}

We used a custom, wireless 64-channel EMG signal acquisition device~\cite{moin2018emg} to record a dataset of EMG signals from five able-bodied, adult male subjects\footnote{Dataset and scripts available at \url{https://github.com/flexemg/flexemg_v2}}. A flexible 16x4 array of electrodes was wrapped completely around the subject's upper forearm, capturing activity of the extrinsic flexor and extensor muscles involved in finger movements with \SI{1}{\kilo S/s} sampling rate. A single Ag/AgCl electrode is attached to the elbow to provide a reference voltage for all channels. The raw recorded signals were wirelessly transmitted to a base station for offline processing. Additionally, we calculated the mean signal energy across all channels as a measure of contraction effort level, and illustrated the value as a bar graph (Fig.~\ref{fig:bar}) in the graphical user interface (GUI). This served as a visual feedback to the subject in real-time.

For this study, we chose a set of gestures consisting of movements of the thumb, index, and middle fingers to model simple grasping actions (Fig.~\ref{fig:gests}): index finger flexion and extension, middle finger flexion and extension, and thumb flexion and extension as single degree-of-freedom (DOF) gesture subset, and one, two, and fist as multi-DOF subset. For each gesture, we started with a calibration phase during which the subject was asked to perform the gesture with the maximum contraction effort, also known as maximum voluntary contraction (MVC). This value was normalized to map to 100\% in the GUI feedback bar graph (Fig.~\ref{fig:bar}). The subject was asked to target three different effort levels (low effort at 25\%, medium effort at 50\%, and high effort at 75\%) for each gesture, repeating each 5 times. 

\begin{figure}
	\centering
	\includegraphics[width=\columnwidth]{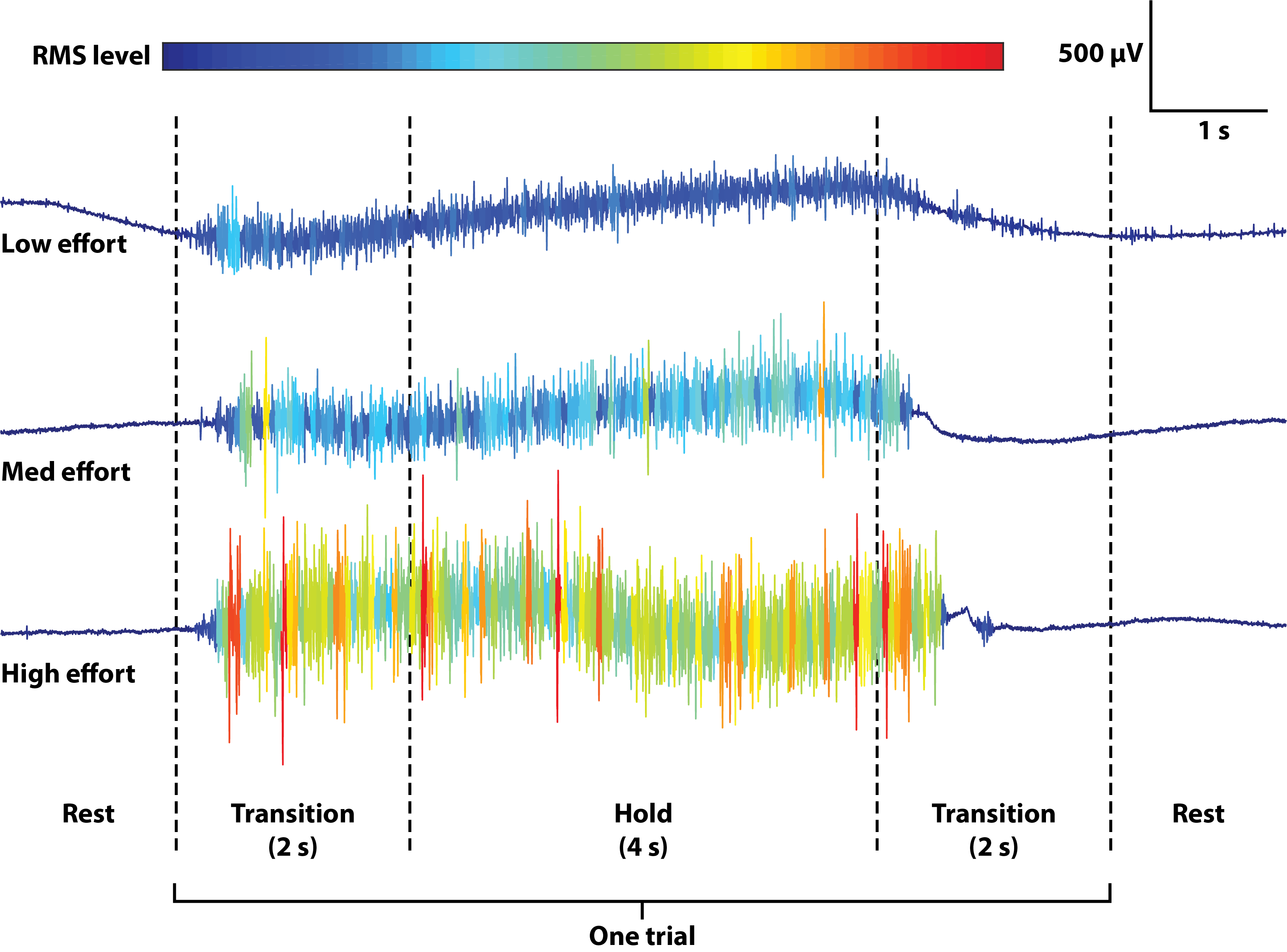}
	\vspace{-20pt}
	\caption{Representative EMG signals from one electrode channel recorded during a low, medium, and high effort level trial of the same gesture. The vertical dotted lines divide a single \SI{8}{\second} gesture trial into \SI{2}{\second} transition periods and a \SI{4}{\second} hold period based on the instructions given to the subject. The color of the waveform indicates the effort level, as measured by windowed signal power (RMS calculated over \SI{200}{\milli\second} windows with \SI{150}{\milli\second} overlap).}
	\label{fig:rawemg}
	\vspace{-10pt}
\end{figure}

Each trial lasted 8 seconds (Fig.~\ref{fig:rawemg}), with 3 seconds of rest before the next trial. The subject was told to begin the gesture within a 2-second transition window which would contain the transient, non-stationary part of the EMG signal for that gesture. After the 2-second transition window, the subject was asked to hold the gesture for 4 seconds, constituting the steady-state part of the EMG signal. Finally, the subject was directed to return to the rest position within another 2-second transition window. These directions ensured that the steady-state portion of the gesture could easily be labeled as part of the middle 4 second segment. Data were automatically labeled with the gesture class and saved as .mat files for processing in MATLAB (MathWorks, Inc.). 

All experiments were performed in strict compliance with the guidelines of IRB and were approved by the Committee for Protection of Human Subjects at University of California, Berkeley (Protocol title: Flex EMG Study. Protocol number: 2017-10-10425).

\section{Classification Algorithm}

\subsection{HD Computing Background}

HD computing employs hypervectors with very high dimensionality (e.g. 10,000) to represent information, analogous to the way the human brain utilizes vast circuits of billions of neurons and synapses~\cite{kanerva2009hyperdimensional}. In general, a fixed symbol table, or item memory (IM), is built from an initial set of HD hypervectors taken randomly from a high-dimensional (e.g. 10,000-dimensional) space. Each hypervector consists of an equal number of randomly placed $+1$'s and $-1$'s. A fundamental property is that, with a very high probability, hypervectors within a randomly generated IM will all be orthogonal to each other, i.e. any pair of hypervectors will differ by approximately 5,000 bits. These hypervectors can be combined to form new composite HD hypervectors using well-defined vector space operations, including point-wise multiplication ($*$), point-wise addition ($+$), scalar multiplication ($\times$), and permutation ($\rho$). Because of the high dimensionality and randomness, HD hypervectors can be combined while preserving the original information.

\begin{figure}
	\centering
	\includegraphics[width=0.6\columnwidth]{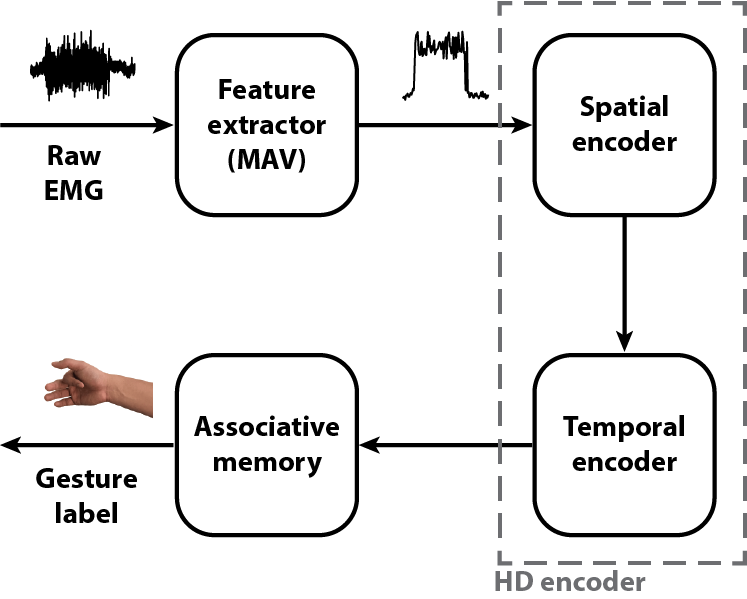}
	\caption{High-level flow diagram for encoding 64 electrode channels of EMG data into hypervectors and outputting the classified gesture label using hyperdimensional (HD) computing algorithm.}
	\label{fig:hd}
	\vspace{-10pt}
\end{figure}

Fig.~\ref{fig:hd} summarizes the process of encoding raw EMG data into HD hypervectors for training and inference. Data is first preprocessed to extract the features to be used as inputs to the HD algorithm. We used mean absolute value (MAV) with non-overlapping windows of 50 samples as input features. Features are then encoded spatially (across 64 channels) and temporally (\SI{250}{\milli\second} windows) into HD hypervectors exactly as described in~\cite{moin2018emg}. Spatiotemporal hypervectors calculated using data from each gesture class are bundled together (i.e., summed) and bipolarized (i.e. positive elements replaced by $+1$ and negative elements replaced by $-1$) to form a binary prototype hypervector representing that class. During training phase, these prototype hypervectors are stored in the associative memory (AM) with their corresponding labels. During inference, the test hypervector is compared to each entry of the AM using cosine similarity as the distance metric. The inferred gesture is selected by finding the closest prototype hypervector in the AM.

\subsection{HD Model for Contraction Effort Levels}

In contrast to many state-of-the-art classification algorithms that often require a big training dataset, HD computing achieves high classification accuracies with small amounts of training data, i.e. only 1 out of 5 trials of each gesture in our case. Therefore, building inclusive prototype hypervectors that contain data from multiple effort levels is fast. If, on the other hand, data from individual effort levels is used to form the prototype hypervectors, HD model will distinguish the effort level in addition to the gesture itself.

\subsubsection{Gesture-Only Classification}

If the only goal is to discriminate between different gestures regardless of the subject's effort level, a single gesture prototype hypervector can be formed to include information from those different effort contexts. This can be done by accumulating spatiotemporal hypervectors from multiple effort levels, and saving its bipolarized hypervector in the AM. If the prototype hypervectors of the two effort levels are already calculated and bipolarized, however, another approach is to merge them into a single prototype hypervector by randomly taking 5000 elements (half of the elements) from each prototype hypervector. 

\subsubsection{Gesture+Effort Classification}

If discriminating between different effort levels of gestures is desired, each \{gesture,effort\} pair must be treated as a separate output class. In this case, prototype hypervectors for each effort level can be added to the model as new entries in AM. 

Note that a potential third case could involve adding new prototype hypervectors for each effort level to the AM while preserving the number of gesture classes, allowing multiple prototype entries to represent the same class. While this will improve the classification accuracy comparing to the case where prototype hypervectors were merged, it costs more memory and computation resources as three prototype hypervectors have to be generated and stored for each gesture class.

\section{Results}

\begin{figure*}
  \centering
  \includegraphics[width=\textwidth]{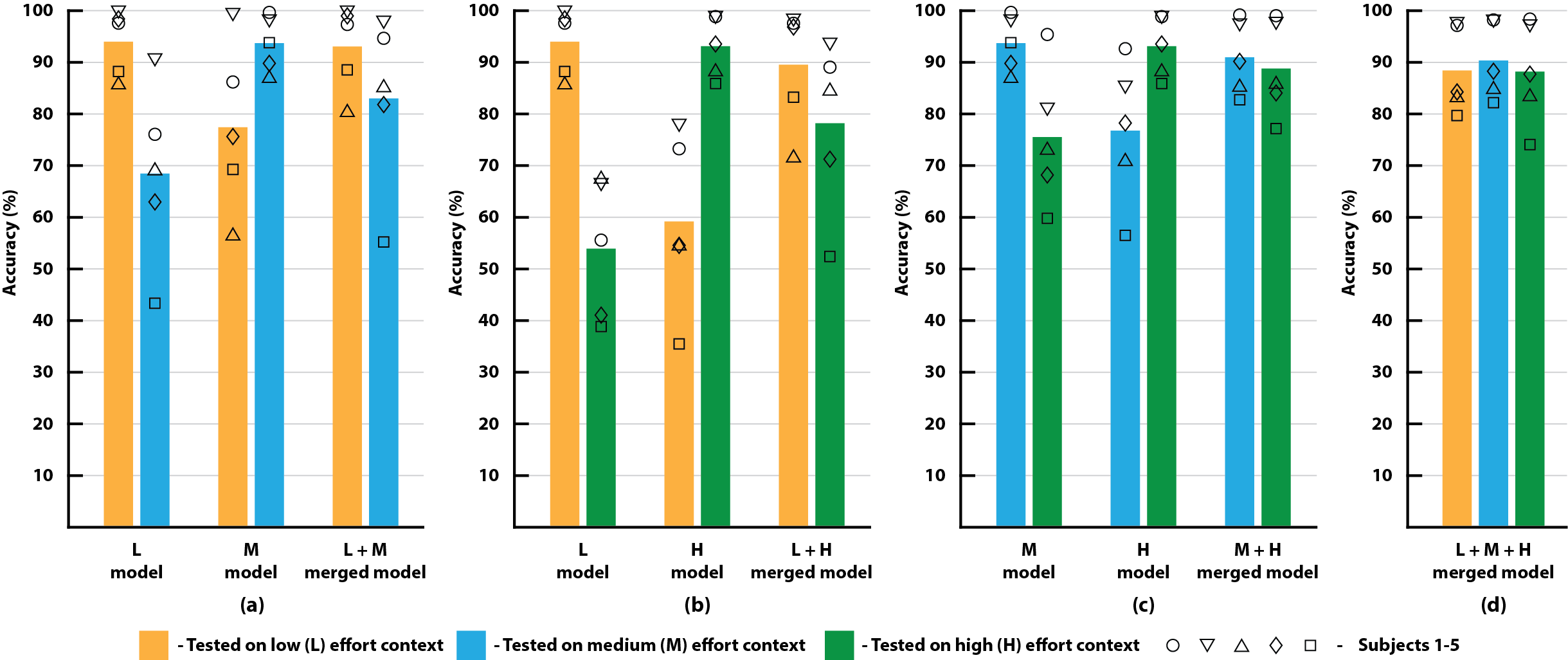}
  \vspace{-5pt}
  \caption{Classification accuracy measured before and after merging the models when treating effort levels as different contexts of the same gesture class. Accuracies across effort contexts before and after merging were calculated for each pair of effort levels: low (L) and medium (M) in (a), low and high (H) in (b), and medium and high in (c). Accuracy for each effort level was also calculated using a model trained with all three effort level contexts (d).}
  \label{fig:samegest}
  \vspace{-5pt}
\end{figure*}

We first treated different effort levels as different contexts of the same gesture class. An initial model was trained with gestures from one effort level context. It was then cross validated (training with one trial, inference with remaining four trials) within the same effort level and also used to classify gestures from the other level within the pair, without merging the models (Fig.~\ref{fig:samegest}(a-c), first and second pairs). When training and testing within the same effort level context, classification accuracy remained better than 93.11\%. However, across different effort levels, classification accuracy dropped by between 16.57\% and 39.17\%, with the worst performance when the difference between effort levels was highest, i.e. low and high effort. After merging the prototype hypervectors to include both effort contexts, the classification accuracy was recovered to above 78.21\% (Fig.~\ref{fig:samegest}(a-c), third pairs). The poorest recovered accuracies resulted from training an initial model on medium or high effort level gestures, and then merging with low effort gestures. While prototype hypervectors for medium and high effort level gestures were more similar to each other, prototype hypervectors for low effort level gestures were more distant due to a smaller variance in the calculated feature values. 

For a model trained with all three contexts by accumulating their spatiotemporal hypervectors before bipolarization, accuracy was at least 88.19\% for all three effort contexts (Fig.~\ref{fig:samegest}(d)). In this case, the all-inclusive final hypervector is weighted to be more similar to medium and high effort level prototype hypervectors enabling higher accuracies in those contexts.

\begin{figure}
	\centering
	\includegraphics[width=\columnwidth]{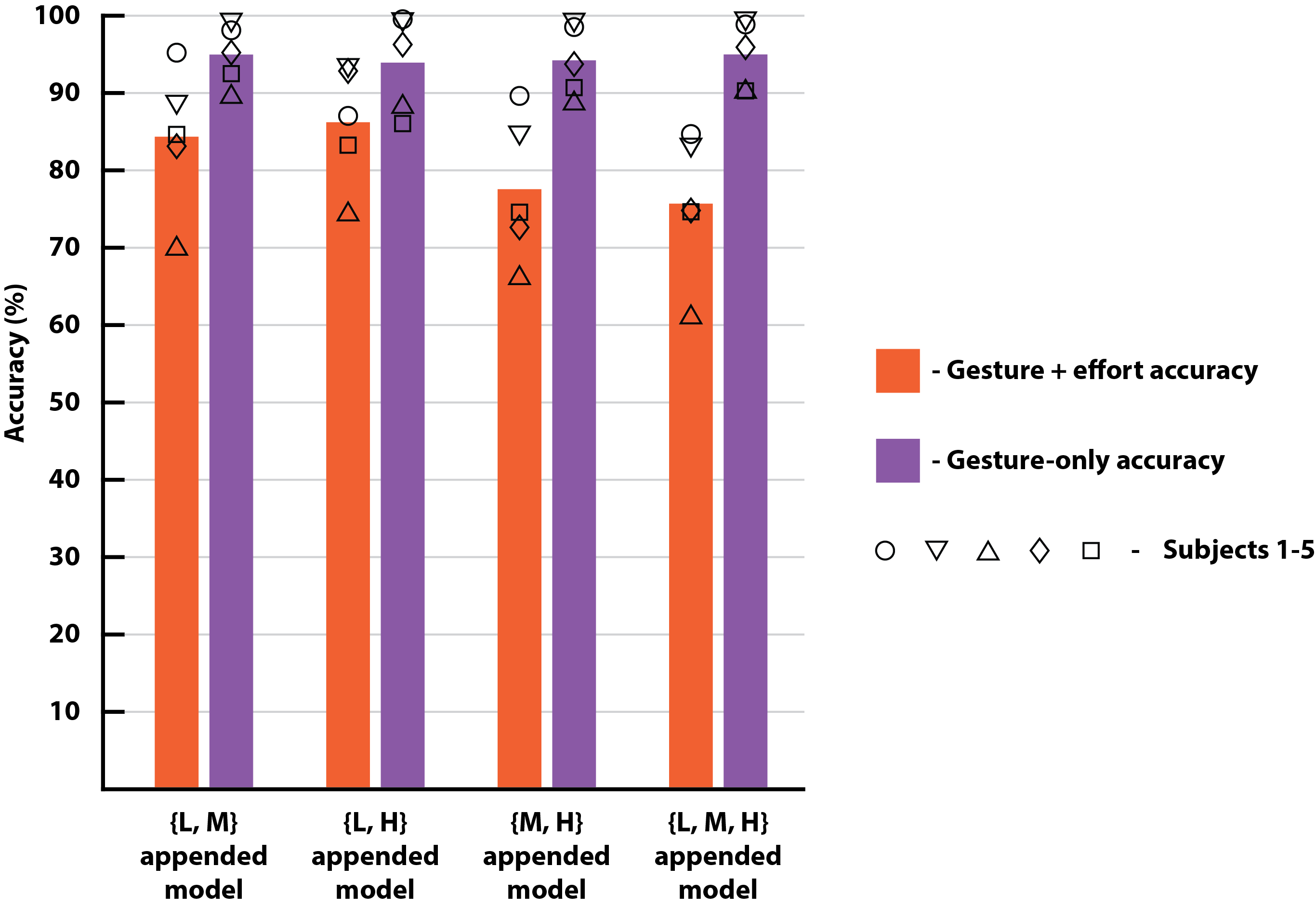}
	\vspace{-5pt}
	\caption{Classification accuracy measured when treating different effort levels of the same gesture as different classes. Accuracy was calculated as the success rate of matching both gesture type and effort level (red) as well as gesture type only (purple).}
	\label{fig:diffgest}
	\vspace{-10pt}
\end{figure}

When treating different effort levels of a single gesture as different classes, we trained a new AM entry for each gesture and effort level, increasing the total number of classes. We calculated classification accuracy in two different ways: For the first method (Fig.~\ref{fig:diffgest}, red bars), an accurate classification required matching both the gesture type and its effort level to the label. For the second (Fig.~\ref{fig:diffgest}, purple bars), an accurate classification required only matching the gesture type. Notably, if we disregard the effort level classification output from this model, we achieve a better gesture-only classification accuracy than in the case where we treated different effort levels as different contexts.

\section{Conclusion}

We have presented methods based on HD computing paradigm that address some of the challenges caused by various muscle contraction levels in EMG-based gesture recognition. Our experimental data showed significant classification accuracy degradation when training and testing across different effort levels. We demonstrated that high accuracy can be simply recovered using a minimum amount of data (only a single trial) from each effort level. Moreover, we verified that the HD model is capable of including new classes to distinguish among multiple effort levels of gestures without the need to change the existing model.

\section*{Acknowledgment}
The authors would like to thank Alisha Menon, George Alexandrov, Senam Tamakloe, Jonathan Ting, Natasha Yamamoto, Yasser Khan, Fred Burghardt, Ken Lutz, Haylie Wu, Profs. Elad Alon, Rikky Muller, Ana C. Arias, Novacentrix Corp. and Cortera Neurotechnologies Inc. This work was supported in part by the CONIX Research Center, one of six centers in JUMP, a Semiconductor Research Corporation (SRC) program sponsored by DARPA. Support was also received from sponsors of Berkeley Wireless Research Center and Savio computational cluster resource provided by the Berkeley Research Computing program at UC Berkeley.

\bibliographystyle{IEEEtran}
\bibliography{IEEEabrv,references.bib}

\begin{thebibliography}{1}
\providecommand{\url}[1]{#1}
\csname url@samestyle\endcsname
\providecommand{\newblock}{\relax}
\providecommand{\bibinfo}[2]{#2}
\providecommand{\BIBentrySTDinterwordspacing}{\spaceskip=0pt\relax}
\providecommand{\BIBentryALTinterwordstretchfactor}{4}
\providecommand{\BIBentryALTinterwordspacing}{\spaceskip=\fontdimen2\font plus
\BIBentryALTinterwordstretchfactor\fontdimen3\font minus
  \fontdimen4\font\relax}
\providecommand{\BIBforeignlanguage}[2]{{%
\expandafter\ifx\csname l@#1\endcsname\relax
\typeout{** WARNING: IEEEtran.bst: No hyphenation pattern has been}%
\typeout{** loaded for the language `#1'. Using the pattern for}%
\typeout{** the default language instead.}%
\else
\language=\csname l@#1\endcsname
\fi
#2}}
\providecommand{\BIBdecl}{\relax}
\BIBdecl

\bibitem{geng2012toward}
Y.~Geng, P.~Zhou, and G.~Li, ``Toward attenuating the impact of arm positions
  on electromyography pattern-recognition based motion classification in
  transradial amputees,'' \emph{Journal of neuroengineering and
  rehabilitation}, vol.~9, no.~1, p.~74, 2012.

\bibitem{young2011improving}
A.~J. Young, L.~J. Hargrove, and T.~A. Kuiken, ``Improving myoelectric pattern
  recognition robustness to electrode shift by changing interelectrode distance
  and electrode configuration,'' \emph{IEEE Transactions on Biomedical
  Engineering}, vol.~59, no.~3, pp. 645--652, 2011.

\bibitem{al2015improving}
A.~H. Al-Timemy, R.~N. Khushaba, G.~Bugmann, and J.~Escudero, ``Improving the
  performance against force variation of {EMG} controlled multifunctional
  upper-limb prostheses for transradial amputees,'' \emph{IEEE Trans. on Neural
  Systems and Rehab. Eng.}, vol.~24, no.~6, 2015.

\bibitem{scheme2011electromyogram}
E.~Scheme and K.~Englehart, ``Electromyogram pattern recognition for control of
  powered upper-limb prostheses: state of the art and challenges for clinical
  use.'' \emph{Journal of Rehabilitation Research \& Development}, vol.~48,
  no.~6, 2011.

\bibitem{khushaba2016combined}
R.~N. Khushaba, A.~Al-Timemy, S.~Kodagoda, and K.~Nazarpour, ``Combined
  influence of forearm orientation and muscular contraction on {EMG} pattern
  recognition,'' \emph{Expert Systems with Applications}, vol.~61, pp.
  154--161, 2016.

\bibitem{he2014invariant}
J.~He, D.~Zhang, X.~Sheng, S.~Li, and X.~Zhu, ``Invariant surface {EMG} feature
  against varying contraction level for myoelectric control based on muscle
  coordination,'' \emph{IEEE journal of biomedical and health informatics},
  vol.~19, no.~3, pp. 874--882, 2014.

\bibitem{kanerva2009hyperdimensional}
P.~Kanerva, ``Hyperdimensional computing: An introduction to computing in
  distributed representation with high-dimensional random vectors,''
  \emph{Cognitive computation}, vol.~1, no.~2, pp. 139--159, 2009.

\bibitem{moin2018emg}
A.~Moin, A.~Zhou, A.~Rahimi, S.~Benatti, A.~Menon, S.~Tamakloe, J.~Ting,
  N.~Yamamoto, Y.~Khan, F.~Burghardt \emph{et~al.}, ``An {EMG} gesture
  recognition system with flexible high-density sensors and brain-inspired
  high-dimensional classifier,'' in \emph{2018 IEEE International Symposium on
  Circuits and Systems (ISCAS)}, 2018.

\bibitem{burrello2018one}
A.~Burrello, K.~Schindler, L.~Benini, and A.~Rahimi, ``One-shot learning for
  {iEEG} seizure detection using end-to-end binary operations: Local binary
  patterns with hyperdimensional computing,'' in \emph{2018 IEEE Biomedical
  Circuits and Systems Conference (BioCAS)}, 2018.

\end{thebibliography}

\end{document}